\documentclass[12pt]{article}
\input epsf.sty
\newcommand{\be}{\begin{equation}}
\newcommand{\ee}{\end{equation}}
\newcommand{\ben}{\begin{eqnarray}\displaystyle}
\newcommand{\een}{\end{eqnarray}}

\newcommand{\sectiono}[1]{\section{#1}\setcounter{equation}{0}}

\def\sqr#1#2{{\vcenter{\vbox{\hrule height.#2pt
         \hbox{\vrule width.#2pt height#1pt \kern#1pt
            \vrule width.#2pt}
         \hrule height.#2pt}}}}

\begin{document}

\begin{center}
\large{\bf The Overshoot Problem and Giant Structures   }

\vspace{10mm}

\normalsize{Nissan Itzhaki \\\vspace{2mm}{\em Tel-Aviv University,
Ramat-Aviv, 69978, Israel}\\\vspace{2mm}nitzhaki@post.tau.ac.il}


\end{center}

\vspace{10mm}

\begin{abstract}


Models of small-field inflation often suffer from the overshoot problem.
 A particularly  efficient resolution to the  problem  was proposed recently in the context of string theory. We show that  this resolution predicts the existence of  giant spherically symmetric overdense regions with radius of at least  $110$ Mpc. We argue that if such structures will be found they could offer an  experimental window into string theory.

\end{abstract}

\newpage

\baselineskip=18pt


\newpage
\renewcommand{\theequation}{\thesection.\arabic{equation}}
\sectiono{Introduction }
\bigskip

Recent observations  strongly support a period of inflation in
the early universe \cite{inf}. In fact  many models of inflation, some of
which were quite popular not too long ago, are now ruled out \cite{WMAP5}.
In light of this remarkable progress it is tempting
to ask, perhaps prematurely, what might be the experimental signatures
 of a pre-inflationary period, assuming such a period existed.
This and related questions were addressed by various authors (see e.g. \cite{rog}).
Our goal here is to connect this question with   the overshoot problem
associated with  small-field inflation.

If the total number of e-foldings during inflation, $N_e^{tot}$,
is much larger then the number of e-foldings needed to resolve the
big-bang puzzles, $N_e^{BB}$, then the probability of finding any
pre-inflationary signature in the visible universe is likely to be negligible. The
difficulties  encountered  in realizing in string theory models of
inflation with a large number of e-foldings    perhaps should be viewed
as an indication that $N_e^{tot}$ is not much larger than
$N_e^{BB}$, in which case  pre-inflationary signatures might be
 detectable.

Most stringy models of inflation suggested so far are models of
small-field inflation.
Such models often suffer from the overshoot problem
\cite{overshoot}, which in this context means  that, against the spirit of
inflation, one needs to  tune the initial conditions of the
inflaton for inflation to take place.  This problem received a lot of attention  over the years (mostly in the context of moduli stabilization), and various ways to overcome it were proposed (see e.g. \cite{atp}).
A nice aspect of the overshoot problem associated with  small-field inflation is that any mechanism which resolves it should act mostly right before inflation, and so
 it is possible that it leaves detectable imprints (assuming  $\Delta N _e=N_e^{tot}-N_e^{BB}$
is not too large).

Recently a particularly efficient resolution
  to  the  problem    was suggested  in \cite{ik}.
The resolution  relies on the existence of particles with mass that
depends on the expectation value of the inflaton. These particles
push the inflaton in the opposite direction than the static potential
does ($m_{,\phi}~V_{,\phi}<0$), and slow down the inflaton  at the slow roll region.
In the present paper we  show
that this pre-inflationary scenario
has a distinct  signature: a formation of giant spherically symmetric overdense regions with a radius of at least  $110$ Mpc.
Each  of these particles  present at the beginning of inflation
provides the seed of a single giant spherically symmetric structure. The properties of these  giant structures are fixed by  $m_{,\phi}$.

The paper is organized as follows:  In the next section we review the mechanism  proposed in \cite{ik} to resolve the overshoot problem.
In section 3 we show that this mechanism leads to the formation of overdense regions and study their properties. We argue  that if
$N_e^{tot}$ is not much larger than
$N_e^{BB}$ some of these giant structures  should be found in the visible universe.

\sectiono{A review  of \cite{ik}}

Most models of inflation in string theory are  small-field
models\footnote{As far as we know the only large-field model of
inflation in string theory was proposed recently \cite{eva}.} (for reviews of
 stringy inflation see \cite{R1}).
Namely, the slow roll conditions are satisfied over  a small distance in the field space
of the canonically normalized inflaton.
Roughly speaking the reason  is that in string theory there are severe
constraints on the possible terms that could contribute to the
inflaton potential. Hence it is hard to construct a potential for the inflaton
that satisfies the slow roll
condition over a large  distance in field space.

Typically models of small-field inflation suffer from the overshoot problem.
This problem was first raised in the context of moduli
stabilization in \cite{overshoot} and is very much relevant also for small-field inflation (see e.g \cite{bgw}).
The problem is that a generic initial condition is
not at the region where the slow roll conditions are met, and if we
start away from the slow roll region the inflaton will overshoot
it  without ever being dominated by the potential
energy. This happens because in small-field inflation, by
definition, the slow roll region  is small and  the Hubble
friction does not have enough time to slow down the inflaton. As a result
 the universe does not inflate despite the fact that the
inflaton passes through the slow roll region.

Most of the resolutions proposed to the overshoot problem \cite{atp} are based on the fact that the energy
density of a scalar field dominated by its kinetic energy  scales like $1/a(t)^6$. Thus, almost any other contribution to the energy density, like matter or radiation, does not drop as fast  and eventually  it takes over. As a result the Hubble friction becomes larger and it could potentially slow down the inflaton at the slow roll region. Typically, however,  these kind of mechanisms are more efficient when addressing the overshoot problem in the context of moduli stabilization than in the context of  small-field inflation.

Recently  \cite{ik} a more efficient resolution to the problem was proposed in the context of string theory\footnote{In relation with the overshoot problem associated with  moduli stabilization a similar mechanism   was proposed  in \cite{krs}. }.  It was noticed in \cite{ik} that at least in some
cases there are  particles with masses that depend
on the inflaton and satisfy
 \be\label{ag} m_{,\phi}~V_{,\phi}<0.\ee
 Such particles will change dramatically  the dynamics of the inflaton
 in the following way.
 Much like any other particles these
particles are expected to be produced thermally when the energy
density is high (Fig. 1(a)). Since $
m_{,\phi}~V_{,\phi}<0$
their density will induce an effective potential that pushes the
inflaton in the opposite direction than the static potential does  (Fig.
1(a)). As the universe expands the density of these particles and
 their effective potential grows weak and the inflaton
minimizes the time dependent  potential energy (Fig. 1(b)). Upon entering the slow
roll region the slope of the static potential becomes negligible
and is not able to  balance the potential induced by the particles
(Fig. 1(c)). Hence  the particles get  diluted while the
inflaton stays in the slow roll region (Fig. 1(d)). This  sets up
the initial condition for the inflaton at the  slow
roll region.

This mechanism is rather general and is expected to work in any
model  that includes particles that satisfy $
m_{,\phi}~V_{,\phi}<0$ with a large enough $m_{,\phi}$.
Let us review how this comes about in the stringy  example considered in
\cite{ik}. The setup is of  modular inflation \cite{mi}. We consider ten dimensional string theory
 with  topology
$R^{3,1}\times X$ where $X$ is a six dimensional compact manifold.
We {\it assume} that all moduli but the  volume of $X$ are
fixed and consider the possibility that the inflaton is related
to  the  volume of $X$. The  relation between the
canonically normalized inflaton and $V_X$ is
 \be\label{cn} L=e^{\phi/\sqrt{24}},~~~\mbox{where}~~~L\sim V_X^{1/6} .\ee
\begin{figure}
\begin{picture}(110,120)(0,0)
\vspace{100mm} \hspace{0mm} \mbox{\epsfxsize=37mm
\epsfbox{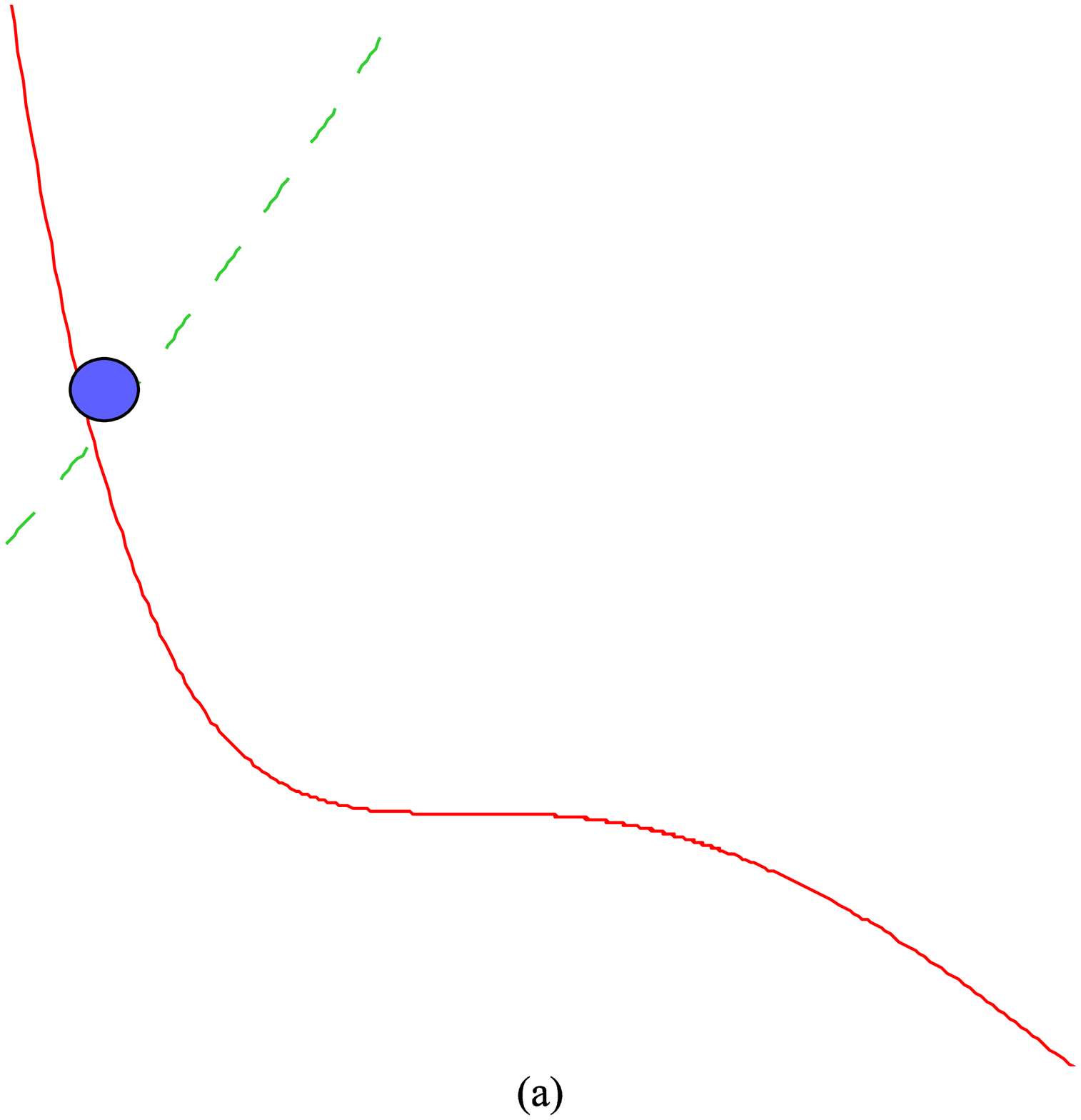}}
\end{picture}
\begin{picture}(110,120)(0,0)
\vspace{0mm} \hspace{0mm} \mbox{\epsfxsize=37mm
\epsfbox{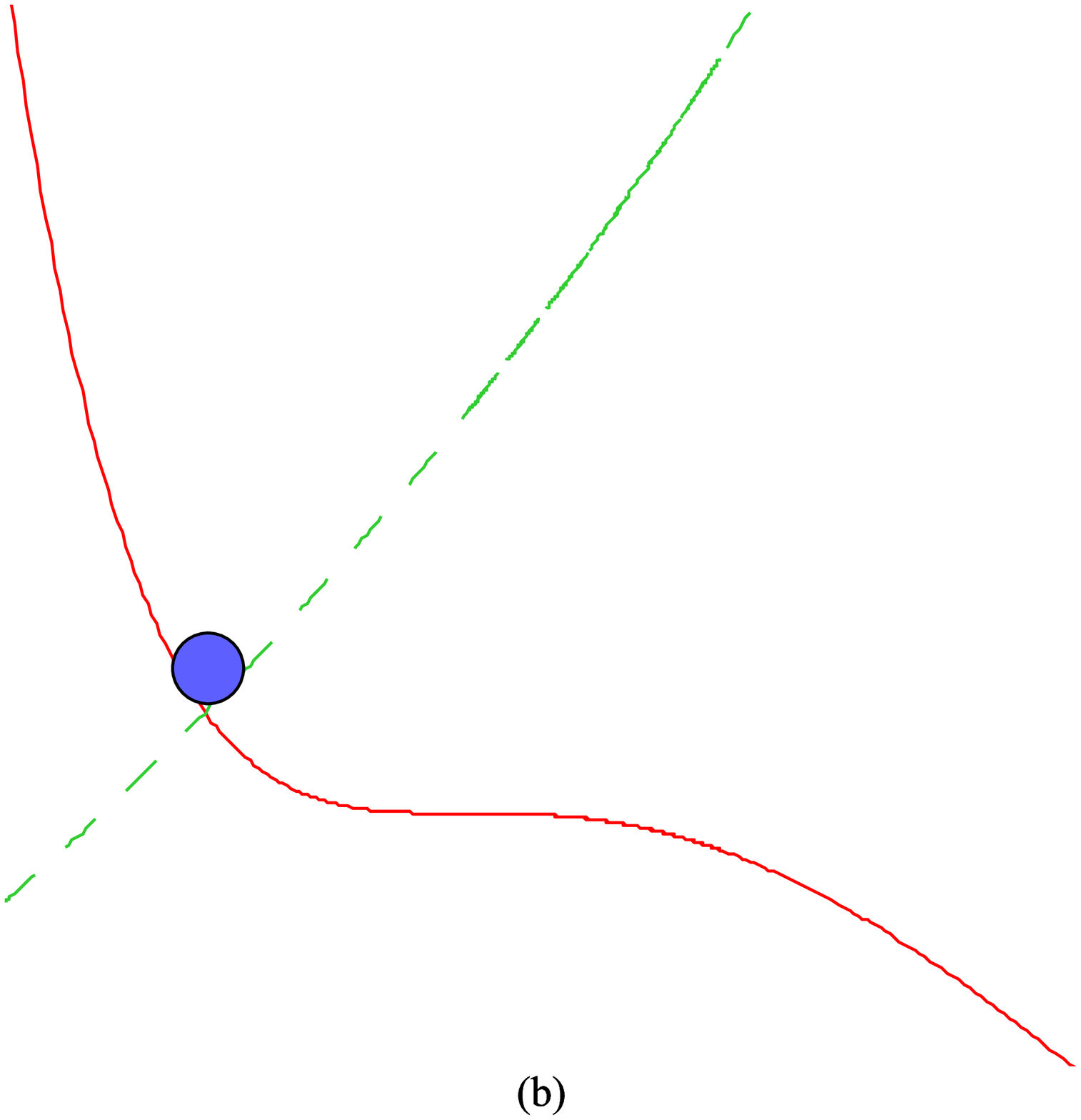}}
\end{picture}
\begin{picture}(110,120)(0,0)
\vspace{0mm} \hspace{0mm}
\mbox{\epsfxsize=37mm\epsfbox{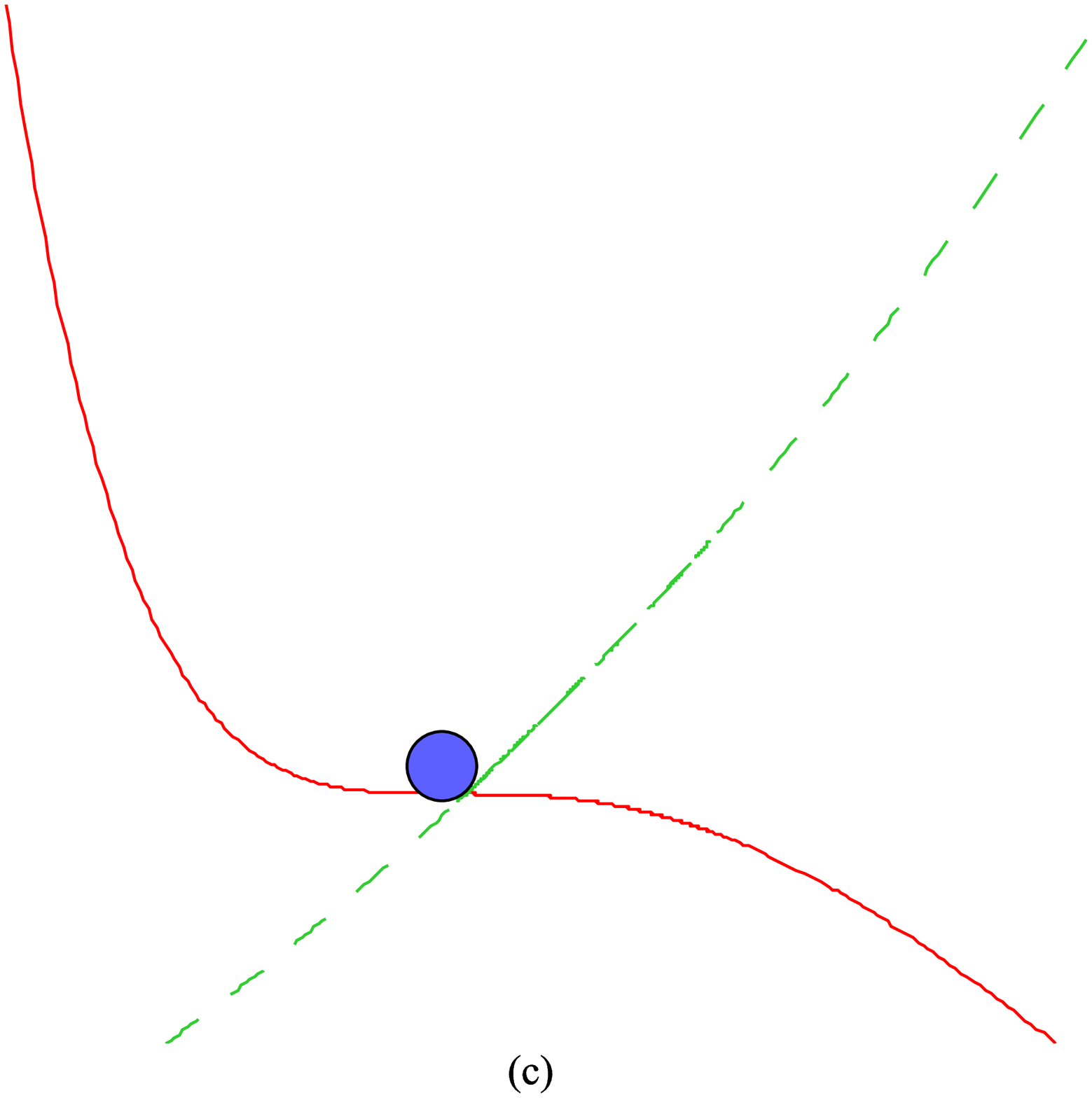} }
\end{picture}
\begin{picture}(110,120)(0,0)
\vspace{10mm} \hspace{0mm} \mbox{\epsfxsize=37mm
\epsfbox{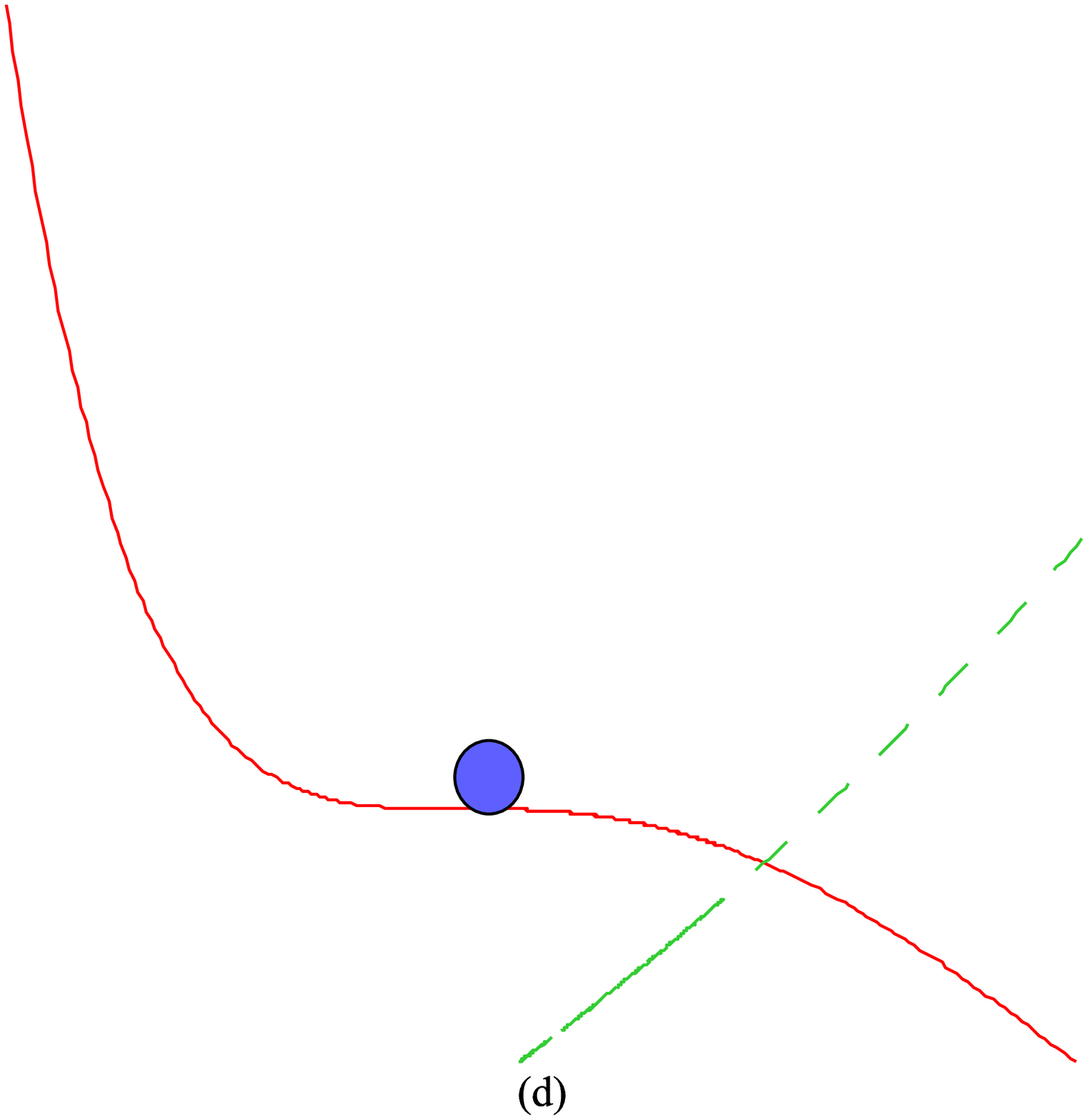}}
\end{picture}
\caption{ A heuristic illustration of how a time dependent
potential, that scales like $1/a^3(t)$, resolves the overshoot
problem of small-field models of inflation. The solid line represents the static potential,  the dashed line the time dependent potential that is induced by the particles and the blue dot the time dependent value of the inflaton.}
\end{figure}
String theory allows only for a small number of terms to appear in
the classical potential for $L$ (or $\phi$). These are due to
wrapped branes, fluxes and curvature. For example in type IIB we
can have
 \be\label{op} V(\phi)=\sum_a C_a L^{-a},~~~~\mbox{with}~~~~a=8,10,12,16, \ee
and in type IIA
 \be V(\phi)=\sum_a C_a L^{-a},~~~~\mbox{with}~~~~a=8,9,10,11,14,18 .\ee
 Some of the constants, $C_a$, can be negative but
most of them cannot, and depending  on the topology of $X$ some of
them have to vanish. Moreover all the terms go to zero at least
as fast as $1/L^8$ when $L\to\infty$. As a result it is
practically impossible  to satisfy the slow roll condition over a
wide region of $\phi$. Namely, in this setup one cannot construct
a large-field model of inflation. In fact it is not easy to
construct a model of small-field inflation  either. The simplest
model  is of an inflection point inflation constructed with the help of
three terms. As argued above and as was
illustrated in  \cite{ik} this model suffers from the overshoot
problem. It can be shown that  the overshoot problem is generic in
this setup.

The nice feature of the model is that it includes also particles
with $ m_{,\phi}~V_{,\phi}<0$ which resolve the overshoot problem via the
mechanism described above. The relevant particles are D-branes
which wrap some of the cycles of $X$. The masses of these
particles are
 \be\label{nper} M_{D-brane}=c_B L^B,~~~~\mbox{with}~~~~B>0,\ee
where $c_B$ is a positive constant that depends on the value of
the other moduli field (in particular the dilaton) which are assumed to be fixed. Since no concrete mechanism for fixing them was proposed in \cite{ik}  $c_B$ cannot be calculated in this approximated setup.
 Again
the possible values of $B$ depend on the type of string theory under consideration, and the topology of $X$. In type IIA the possible values of $B$ are $1$ or $ 3$
and in type IIB there is only one possible value
 $ B=2.$

A natural question to ask is whether these particles lead to any clear experimental prediction
 which could test the mechanism of \cite{ik}.
 Much like monopoles, or any other heavy particles, they will get
diluted exponentially fast  during inflation, and so their contribution to the
total energy in the universe at the end of inflation is negligible. On the other hand
their imprint on structure formation could be  significant and perhaps even detectable. This is the subject of the next section.

\sectiono{Giant  Structures}

The discussion  in the previous section was done in the homogenous approximation. Namely, we considered the net effect  particles with $m_{,\phi}V_{,\phi}<0$ have on the evolution of the zero mode of the inflaton.
During  inflation
the proper distance between the  particles grows exponentially fast and very quickly the
 homogenous approximation breaks down at macroscopic scales. Since  these particles couple {\it directly} to the inflaton and since this coupling played such an important role in  the dynamics of the inflaton we expect  the inhomogeneities due to the individual  particles to induce inhomogeneities of the  inflaton.\footnote{On top of the usual  inflaton's inhomogeneities due to quantum effects.}
 It is well known that inhomogeneities of the  inflaton  provide the seeds of structure formation.
 Hence,  it is reasonable to suspect that these particles could   affect  structure formation in an interesting way.

We wish to
 study the effect
  a single particle with  $m_{,\phi}V_{,\phi}<0$   present at the beginning of inflation has on structure formation.
 Before we turn to the actual  calculation it is instructive   to recall the  intuitive  relation between the inflaton and
structure formation \cite{Guth:1982ec}. During inflation $\phi$ plays the
role of a clock and in particular it determines the time at which inflation
ends. A non-uniform inflaton  will
cause inflation to end in a non-uniform fashion. Regions where
inflation ends first had more time to expand after inflation and
so their density is slightly smaller.
   A particle with $m_{,\phi}V_{,\phi}<0$  pushes the inflaton in the opposite direction than the static potential does. Hence inflation will end first away from the particle. Therefore, we expect such a  particle to provide the seed of an overdense region.

 To estimate the size and density of this overdense region we start by calculating the non-uniform shape of the inflaton caused by  a single particle with $m_{,\phi}V_{,\phi}<0$.
 For simplicity we consider a spatially  flat  universe
 $$ ds^2=-dt^2+a(t)^2 dx_i^2.$$
Since the mass of the particle depends on the inflaton its presence   modifies
 the inflaton
equation of motion  directly
 \be\label{basic} \ddot{\phi}+3H \dot{\phi} - \frac{1}{a(t)^2}\nabla^2 \phi
 + V_{,\phi} +m_{,\phi} \frac{\delta^3(x_i)}{a(t)^3}=0.\ee
To solve this equation we separate the
solution into two parts: the zero mode and the $r$ dependent perturbation
 \be \phi(r,t)=\phi(t) + \delta\phi(r,t). \ee
 $\phi(t)$ solves the standard equation
 \be\label{zero_mode} \ddot{\phi}+3H
 \dot{\phi} + V_{,\phi}=0 ,\ee
and $\delta\phi(r,t) $ solves the linear equation
 \be \ddot{\delta\phi}+3H \dot{\delta\phi} -
 \frac{1}{a(t)^2}\nabla^2
\delta\phi + V_{,\phi\phi}\delta\phi +
 \frac{m_{,\phi}}{a(t)^3} \delta^3(x_i)=0.\ee
During the period of slow roll inflation the $V_{,\phi\phi}$ term
can be neglected and we have
 \be\label{master} \ddot{\delta\phi}+3H
\dot{\delta\phi} - \frac{1}{a(t)^2}\nabla^2 \delta\phi  +
 \frac{m_{,\phi}}{a(t)^3} \delta^3(x_i)=0.\ee
As expected we ended  up with  the familiar  equation for the perturbation
only that now there is a source located at the origin.

At early times, when $r a(t) \ll 1/H$, the time derivatives in
(\ref{master}) are negligible and the solution takes the familiar
form
 \be\label{icon} \delta\phi=-\frac{m_{,\phi}}{4\pi r a(t)},\ee
 which in momentum space gives
 \be\label{ll} \delta\phi_k=- \frac{m_{,\phi}}{(2\pi)^{3/2}k^2 a(t)}
 .\ee
At late times the Hubble friction term  in
(\ref{master}) becomes
important and  cause a freeze out of different
 modes at different times. As usual in an accelerating universe the freeze out occurs
when $a(t)\sim k/H$
and so  the late time solution reads
 \be\label{df}  \delta\phi_k=-C\frac{H m_{,\phi}}{k^3},\ee
where $C$ is a positive constant.  In the
   appendix we calculate $C$ and find
 \be\label{C} C=\frac{1}{\sqrt{32\pi}}.\ee

Next we use the well know  relation (see e.g. \cite{ll}) between the  inflaton  and structure formation
\be\label{td} \delta_k=\frac25 \frac{k^2}{H_0^2} T(k) R_k.\ee
Here  $\delta_k$ is the momentum mode of $\delta\rho/\rho$, $H_0$ is the present Hubble scale,  $R_k$ is determined during inflation
\be R_k=-\frac{H_{inflation}}{\dot{\phi}}\delta\phi_k,\ee
and $T(k)$ is the transfer function.

For simplicity we make two approximations. First we take the most naive
approximation to the transfer function
\ben  T(k) = \left\{\begin{array}{ll}
1 & \mbox{for}~~ k< k_{eq}\\\label{fed}
k_{eq}^2/k^2 & \mbox{for}~~ k> k_{eq},  \end{array}\right. \een
 where $1/k_{eq}$ is the comoving  Hubble length at matter-radiation
equality $\sim 14 \Omega^{-1}_M h^{-2} \sim 110$ Mpc.
This approximation captures, in a crude fashion,  the fact that the formation of structure during  the radiation dominated era is suppressed compered to the matter dominated era. However it neglects the less clean physics associated with modes which enter the horizon during the radiation dominated era. The corrections to (\ref{fed})  are logarithmic in $k$  (see e.g. \cite{ll}).

The second approximation is  the fact that we ignore  the effects due to dark energy which become important at late times. That is  we take a trivial growth function.  If the size of the formed structure  is not too
large a significant portion of the evolution is in the matter
dominated era, and the approximation is expected to be fairly good.

Clearly these approximations are rather crude, and
to have a better description of the formed structure we need to go beyond them. This will be done elsewhere. Here we take advantage of the fact that these approximations  yield simple expressions that, we believe, capture the main effects.

Let us first assume that the  modes that are most relevant for the structure formation are modes with $k<k_{eq}$. Momentarily we will see what is the condition   $m_{,\phi}$ has to satisfy  for this to be the case.  In this case  $T(k)=1$, and
\be \delta_k=-\frac25 \frac{k^2}{H_0^2}\frac{H_{inflation}}{\dot{\phi}}\delta\phi_k.\ee
Using the slow roll equation  ($3H \dot{\phi}=-V_{,\phi}$) and  (\ref{df}, \ref{C}) we get
\be \delta_k=-\frac{V^{3/2}m_{,\phi}}{10\sqrt{6\pi}H_0^2 V_{,\phi} k}=1.2\times 10^{-5} \frac{|m_{,\phi}|}{H_0^2 k},\ee
where in the second equality we  used the COBE normalization  ($ \frac{V^{3/2}}{V_{, \phi}}=5.2\times 10^{-4}$)
and (\ref{ag}).

Transforming back into position space we find that
 \be \frac{\delta\rho}{\rho}\sim 10^{-5} \frac{|m_{,\phi}|}{H_0^2 r^2},\ee
and that the size of the overdense region  (fixed by $\delta\rho=\rho$) is
 \be\label{size} r\sim
 m_{,\phi}^{1/2} 13 ~\mbox{Mpc}.\ee
We see that for the assumption that the most relevant modes  are modes with  $k<k_{eq}$ to hold $m_{,\phi}$ should be larger than $\sim (110/13)^2\sim 70$, which is quite large given that  $m_{,\phi}$  is a dimensionless parameter.

Next we consider the probably more realistic  case with $m_{,\phi}<70$ in which the relevant modes have $k>k_{eq}$. Then  $T(k)=k_{eq}^2/k^2$ and
\be \delta_k=-\frac25 \frac{k_{eq}^2}{H_0^2}\frac{H_{inflation}}{\dot{\phi}}\delta\phi_k.\ee
Following the same steps as before we get
\be \delta_k=1.2\times 10^{-5} \frac{k_{eq}^2|m_{,\phi}|}{H_0^2 k^3}=1.6\times 10^{-2} \frac{|m_{,\phi}|}{ k^3}.\ee
Transforming back into position space we find  for $r<k^{-1}_{eq}\sim110$ Mpc that $\frac{\delta\rho}{\rho}$ is approximately a constant\footnote{Neglecting logarithmic effects which are ignored since the transfer function we work with, (\ref{fed}),  neglects various logarithmic corrections. }
\be\label{yg}  \frac{\delta\rho}{\rho}\sim   10^{-2}  |m_{,\phi}|.\ee
Therefore, for $m_{,\phi}$ of order $1$ we expect $\frac{\delta\rho}{\rho}$ to be fairly small $\sim 10^{-2}$. However, this small effect takes place within  a giant spherically symmetric region with a radius of about  $110$ Mpc.
Within this region we expect to find the usual  structures  (due to  quantum fluctuations during inflation) at much smaller scales but with $\frac{\delta\rho}{\rho}$ much larger than $ 10^{-2}  |m_{,\phi}|$.

Put differently each particle  creates a giant region  of size of  order $220$ Mpc in which the universe looks roughly like it used to look at $z\sim m_{,\phi}/100$. That is, it has similar structure as in the rest of the universe:
Most of the galaxies are in a nearly spherically symmetric  overdense region of size of a few Mpc.
Some of these almost spherically symmetric regions  are connected via filaments with smaller density and size  of order 10-30 Mpc. The filaments are connect via walls with an even lower density, and there are voids between the walls.
The only difference is that the average density is larger than in the rest of the universe.

 Natural questions to ask at this stage are:\\
1. How large   is $m_{,\phi}$ in the scenario of \cite{ik}?\\
2. How many such overdense regions should we expect to find in the visible
universe?

Let us begin with the first question. As described in the previous section quite generically we expect $m_{,\phi}$  to be fairly large in order to resolve the overshoot problem.
From (\ref{cn}, \ref{nper}) we find that in the concrete example studied in  \cite{ik}
\be\label{re} m_{,\phi}=\frac{c_B ~B}{\sqrt{24}} L_{SR}^B.\ee
In \cite{ik} it was shown that the COBE normalization implies
\be\label{er} L_{SR} \cong 50 a_3^{1/8} ,\ee
where $a_3$ is a constant that, much like $c_B$, depends on the
values at which the other moduli are fixed and so it is not known
within the approximated setup of \cite{ik}.
Combining (\ref{re}) with (\ref{er}) we get
\be\label{la} m_{,\phi} \cong c_B B a_3^{B/8} \frac{ 50^B}{\sqrt{24}} .\ee
We see that unless $c_B$ and $a_3$ obtain unusually small values   $m_{,\phi}$  is  larger then $1$. Hence  the effect  such a particle has on structure formation should be significant.

It is less clear  whether $m_{,\phi}$ is smaller or larger than $\sim 70$. Again with the assumption that $c_B$ and $a_3$ are of order one we see that for   $B=1$ it is most likely that $m_{,\phi}<70$ while for
$B=3$ we probably have $m_{,\phi}>70$. The type IIB case ($B=2$) tends to give   $m_{,\phi}>70$, but it could easily give $m_{,\phi}<70$.
Thus we do not know if such a particle will lead to the formation of a structure larger than $110$  Mpc or not.
We reemphasis that in models in which all modui but the  inflaton are fixed we do expect to be able to compute $m_{,\phi}$ and to make  sharp prediction about the giant structure.

We assumed so far that the particles are isolated. However
since they interact also via a scalar (the inflaton) exchange it is possible that they form clusters in the pre-inflationary stage.\footnote{We thank the referee of the paper for raising  this issue.}
In \cite{663} this  was shown to happen  in the mass varying neutrinos scenario of \cite{800}. Such a cluster of particles will have  $m_{,\phi}^{eff}=N_{cluster}~ m_{,\phi}$. Thus  this could drastically improve the local efficiency of the mechanism of \cite{ik} especially  when $m_{,\phi}$ is small (which is likely to be the case in an effective field theory set-up, since $m_{,\phi}$ is dimensionless) and could change the properties of the giant structure.
This interesting possibility will be studied elsewhere.

 Unfortunately, even in models  in which all moduli but the inflaton are fixed, the answer to the second question is not clear. The reason is that
the answer  is exponentially sensitive  to $N_e^{tot}$:
The density of these particles
at the beginning of inflation, $n_0$, must be large enough to slow
down the inflaton at the slow roll region.  A rough lower bound on $n_0$  comes from demanding   that  just before  infaltion begins
the slope of the  potential induced by  the particles
is larger then the one of the static potential. This gives
 \be n_0 >\frac{V_{SL}}{m_{,\phi}}.\ee
Using the COBE normalization\footnote{Note that  the COBE normalization  fixes $V^{3/2}/V_{, \phi}~$ at around $N_e^{BB}$ e-foldings before the end of inflation, while here we are using it $N_e^{tot}$ e-foldings before the end of inflation. The difference between the two is not necessarily negligible. However, since we are merely trying to estimate $N_0$ this is a reasonable approximation to use.}  we find that the total number of
particles within the Horizon at the beginning of inflation is
 \be N_0 =\frac{n_0}{H^3}>\frac{1}{m_{,\phi} H}\sim
 \frac{10^4}{m_{,\phi}\sqrt{\epsilon}}.\ee
Typically in small-field inflation  $\epsilon\ll \eta \sim 0.05$. Thus
a fair  estimate is  $N_0 \sim 10^6$. This implies that the number of
giant overdense regions  in the visible universe is about
 \be 10^6 e^{ -3(N_e^{tot}-N_e^{BB})},\ee
and that for the number of giant overdense regions in the universe to be of order
$1$ we should have
 \be\label{es} \Delta N_e =N_e^{tot}-N_e^{BB}\sim 5.\ee
With our present  knowledge it is hard
to tell whether this is likely or not to be the case. An argument against this is
that $\Delta N_e/ N_e^{BB} \sim 1/10$. So there is an extra tuning
in the model. On the other hand one can argue that since the
amount of fine tuning needed in order to have $N_e^{tot}\gg 1$
grows with $N_e^{tot}$  it is likely that $N_e^{tot}$ is
not much larger then  $N_e^{BB}$.

We emphasis that  (\ref{es}) is merely
an estimate based on the assumption of high scale inflation. In
low scale inflation  with $V_{SL} $ as low as $\mbox{TeV} ^4$  \cite{lsi} we find that for the number of giant overdense regions in the universe to be of order
$1$ we should have
 \be\label{ess} \Delta N_e \sim 10,\ee
and, since in low scale inflation $N_{e}^{BB}\sim 40$, that   $ \Delta N_e/ N_e^{BB} \sim 1/4$.

Note that there are other particles in the model with mass that depends on the inflaton, but with $ m_{,\phi}~V_{,\phi}>0$. These are the perturbative excitations
with
\be\label{per} m_{per}\sim L^{-A},~~A>0,\ee
which are also expected to be produce thermally at stage (a) of Fig. 1. Hence   some of them are expected to be found at
 the begging of inflation.
Should we conclude from this that if giant overdense regions  are found then giant voids  should be found as well? At least in the model of \cite{ik} the answer is no. The reason is that since from (\ref{er}) we expect $L_{SR}\gg 1$ we find from (\ref{per}) that for the perturbative excitation
 \be |m_{,\phi}^{per}|\ll 1. \ee
Hence the effect of these particles on structure formation is expected to be
negligible (even when $\Delta N_e$ is small).  In fact, this condition must be satisfied. Otherwise  the mechanism described in section 2 will not  resolve the overshoot problem as the induced potential due to  the particles with  $ m_{,\phi}~V_{,\phi}>0$ will cancel the induced potential due to  the particles with  $ m_{,\phi}~V_{,\phi}<0$.

This of course does not mean that there are no other models in which giant voids could be formed via the mechanism described here (with $ m_{,\phi}~V_{,\phi}>0$). We mention this since, in   relation with the WMAP cold spot \cite{cs}, some arguments were
already made for a giant void  \cite{Inoue:2006fn}. See however \cite{Smith:2008tc}.

\sectiono{Summary}

In this paper we showed that the pre-inflationary scenario proposed in
\cite{ik}  to resolve the overshoot problem of small-field
inflation leads to the formation of  giant spherically symmetric overdense regions.  The number of these giant overdense regions  in the visible universe
is exponentially sensitive to $\Delta N_e$, and so  cannot be determined with our present knowledge.

 Since  typically the
structure in the universe is not formed in a spherically symmetric fashion
 this appears to be  a distinct feature of \cite{ik} that   cannot be confused with other possible imprints  due to  finite $\Delta N_e$ or higher order terms.
Hence
we believe that even a detection of a single giant spherically symmetric  overdense region with a radius of at least  $110$ Mpc should be viewed as evidence for the scenario of \cite{ik}. What in our opinion  should be viewed as a  clear cut  evidence for  \cite{ik} is   a detection of several spherically symmetric giant overdense regions with the same properties. The reason is that  it seems extremely unlikely that a different  scenario  could lead to a similar anomaly in structure formation.

Since  the properties of the giant structure are fixed by  a specific parameter in the theory, $m_{,\phi}$,  such a development could perhaps open an interesting dialog
 between cosmology and string theory.

\vspace{10mm}

\noindent {\bf Acknowledgements}

\vspace{4mm}

I thank V. Acquaviva, R. Brustein, E. Koevtz  and especially A. Fialkov for discussions. This work is supported in part by the Marie Curie Actions under grant
MIRG-CT-2007-046423.

\appendix
\sectiono{Appendix}

Here we derive (\ref{C}). The starting point is (\ref{master}).
To solve this equation it is useful to define the
Sasaki-Mukhanov variable \cite{sm}
 \be v \equiv a \delta \phi\ee
and to switch to conformal time, $d\tau=dt/a(t)$, which in de-Sitter space gives
 \be \tau=-\frac{1}{Ha}.\ee
In these variables  (\ref{master}) reads
 \be \ddot{v}-\left( \nabla^2 +\frac{\ddot{a}}{a} \right) v
 =- m_{,\phi} \delta^3(x_i),\ee
which in momentum space gives
 \be\label{mom} \ddot{v_k}+\left( k^2 -\frac{\ddot{a}}{a} \right) v_k
 =- \frac{m_{,\phi}}{(2\pi)^{3/2}} .\ee
In de-Sitter space $\frac{\ddot{a}}{a}=\frac{2}{\tau^2}$ and the
homogenous solutions ($m_{,\phi} =0$) take the form
 \be v_k= A_1(k) F_1(k,\tau) + A_2(k)  F_2(k, \tau)
, \ee where
 \be F_1(k,\tau)= \frac{1}{\tau}\left( \cos(\tau k) \tau k-\sin(\tau k)
 \right), ~~~~F_2(k,\tau)=\frac{1}{\tau}\left( \cos(\tau k) +\sin(\tau k)\tau k
 \right).\ee
 These solutions  have the well known behavior. At
early times when $\tau k \gg 1$ they oscillate
 \be\label{ic} F_1(k,\tau) = k\cos(\tau k), ~~~~F_2(k,\tau)= k\sin(\tau k),\ee
and at late times ($\tau\to 0^{-}$) they give a decaying mode and a growing mode
(that becomes a constant when transforming back to the original
$\delta \phi_k$)
 \be\label{late} F_1(k,\tau)= -\frac13 \tau^2 k^3,~~~~F_2(k,\tau)=
 \frac{1}{\tau}.\ee
The inhomogeneous solution of
 \be\label{mom} \ddot{v_k}+\left( k^2 -\frac{2}{\tau^2} \right) v_k
 = g(t) \ee
can be written in the following  form
 \be\label{solgen} v_k(\tau)= \frac{1}{k^3}\left( F_2(k,\tau) \int^{\tau}
F_1(k,\tilde{\tau}) g(\tilde{\tau})d\tilde{\tau}- F_1(k,\tau)
\int^{\tau} F_2(k,\tilde{\tau})
 g(\tilde{\tau})d\tilde{\tau}\right). \ee
We are interested in the case $g(t)=- \frac{m_{,\phi}}{(2\pi)^{3/2}} $ in which
 \ben \int^{\tau}
F_1(k,\tilde{\tau}) g(\tilde{\tau})d\tilde{\tau}&=&- \frac{m_{,\phi}}{(2\pi)^{3/2}}
\left(
\sin(\tau k) -\mbox{Si}(\tau k) +C_1\right),\\
 \int^{\tau}
F_2(k,\tilde{\tau}) g(\tilde{\tau})d\tilde{\tau}&=&- \frac{m_{,\phi}}{(2\pi)^{3/2}} \left(
 -\cos(\tau k)+\mbox{Ci}(\tau k) +C_2\right). \een
The constants of integration $C_1$ and $C_2$ are fixed by  the
initial condition (\ref{icon}) at $\tau\to -\infty$, which
in terms of the variable $v_k$ is
 \be\label{ll} v_k=- \frac{m_{,\phi}}{(2\pi)^{3/2}k^2}
 .\ee
This implies that
 \be C_1=\mbox{Si}(-\infty)=-\frac{\pi}{2}, ~~~~~C_2=-\mbox{Ci}(-\infty)=0.\ee
Therefore at late times ($\tau\to 0^{-}$) we find that
 \be v_k=\frac{ m_{,\phi}}{ \sqrt{32 \pi}\tau k^3},\ee
and that
 \be \delta \phi_k =-\frac{ m_{,\phi} H}{ \sqrt{32 \pi} k^3}.\ee


\end{document}